\documentclass[aps,epsf]{revtex4}
\usepackage{amssymb,amsmath}
\usepackage{graphics}
\usepackage{graphicx}
\usepackage{epsfig}

\def\beq{\begin{equation}}
\def\eeq{\end{equation}}
\def\bea{\begin{eqnarray}}
\def\eea{\end{eqnarray}}

\begin{document}

\input epsf.tex

\title{Bayesian Inference for LISA Pathfinder using Markov Chain Monte Carlo Methods.}

\author{Luigi Ferraioli}
\email{luigi.ferraioli@apc.univ-paris7.fr}
\affiliation{APC, Univ. Paris Diderot, CNRS/IN2P3, CEA/Ifru, Obs de Paris, Sorbonne Paris Cit\'e, France}

\author{Edward K. Porter}
\affiliation{APC, Univ. Paris Diderot, CNRS/IN2P3, CEA/Ifru, Obs de Paris, Sorbonne Paris Cit\'e, France}

\author{Eric Plagnol}
\affiliation{APC, Univ. Paris Diderot, CNRS/IN2P3, CEA/Ifru, Obs de Paris, Sorbonne Paris Cit\'e, France}

\begin{abstract}

We present a parameter estimation procedure based on a Bayesian framework by applying a Markov Chain Monte Carlo algorithm to the calibration of the dynamical parameters of a space based gravitational wave detector. The method is based on the Metropolis-Hastings algorithm and a two-stage annealing treatment in order to ensure an effective exploration of the parameter space at the beginning of the chain. We compare two versions of the algorithm with an application to a LISA Pathfinder data analysis problem. The two algorithms share the same heating strategy but with one moving in coordinate directions using proposals from a multivariate Gaussian distribution, while the other uses the natural logarithm of some parameters and proposes jumps in the eigen-space of the Fisher Information matrix. The algorithm proposing jumps in the eigen-space of the Fisher Information matrix demonstrates a higher acceptance rate and a slightly better convergence towards the equilibrium parameter distributions in the application to LISA Pathfinder data .  For this experiment, we return parameter values that are all within $\sim1\sigma$ of the injected values.  When we analyse the accuracy of our parameter estimation in terms of the effect they have on the force-per-unit test mass noise estimate, we find that the induced errors are three orders of magnitude less than the expected experimental uncertainty in the power spectral density.

\end{abstract}



\maketitle

\section{Introduction}
\label{intro}
LISA Pathfinder~\cite{Armano2009,Antonucci2011a,McNamara2008} is a European Space Agency mission aiming to develop and test the key technology for the forthcoming space-based gravitational wave detectors ~\cite{AmaroSeoane2012}. A considerable amount of the LISA Pathfinder mission time will be dedicated to a series of experiments for the calibration of the model parameters for the spacecraft dynamics. Such procedures will be periodically replicated on future space-based gravitational wave detectors in order to ensure that the sensitivity of the instrument stays at the values required for the measurement of gravitational waves. Since the instrument calibration has a direct impact on the scientific outcome of the mission, it is of fundamental importance to develop a powerful and solid procedure for parameter estimation that allows an accurate calibration in a short amount of time. 

The problem of the calibration of parameters describing the dynamics was already discussed in previous works  focusing on LISA Pathfinder~\cite{Antonucci2011b,Congedo2012,Nofrarias2010}. In Ref~\cite{Antonucci2011b,Congedo2012} a frequentist approach to the problem was developed using both linear and non-linear parameter estimation techniques respectively. In Ref~\cite{Nofrarias2010} a Bayesian framework was developed for the second  LISA Pathfinder mock data challenge,  based on a classic Metropolis-Hastings Markov Chain Monte Carlo (MCMC) scheme for parameter estimation\cite{Metropolis1953,Hastings1970}. From the point of view of the present paper, the method presented in Ref~\cite{Nofrarias2010} is comparable with the algorithm working with the direct physical parameters and proposing the MCMC jumps from a multivariate Gaussian distribution.

The MCMC algorithm is a stochastic method that is very useful for carrying out Bayesian inference.  The algorithm allows us to easily sample from a sometimes unknown probability distribution.  By generating an ergodic Markov chain, we can sample from the underlying target distribution.  The longer the MCMC, the more statistically independent the chain is and the better we model this distribution.  In general MCMC algorithms are relatively easy to implement.  While guaranteed to converge to the target distribution, the convergence time of the MCMC is problem dependent and is virtually impossible to estimate. However, the convergence can be accelerated if we use proposal distributions for the chain which we believe to be close to the target distribution.  MCMC algorithms are normally divided into two parts : a ``burn-in" phase where we allow the chain to find an equilibrium state, and a true MCMC which we use afterwards for statistical purposes.  The end of the burn-in phase is usually decided ``by-eye" as it is problem specific.

We propose a Bayesian approach to the calibration of the parameters of the spacecraft dynamics that is based on Metropolis-Hastings MCMC. The  Metropolis-Hastings MCMC is a reference method for parameter estimation in Bayesian Statistics (see for example Ref~\cite{Hitchcock2003}) and in space-based Gravitational waves data analysis (for an overview, we refer the reader to Ref~\cite{Porter2009} and references therein) where it has demonstrated its flexibility and power in the estimation of parameters for a variety of GW sources~\cite{Babak2008} .   We compare here two different algorithms for the construction of the Markov Chain. One works with the physical parameters and proposes MCMC jumps from a multivariate Gaussian distribution, the other uses a conversion from some of the physical parameters to their natural logarithm in order to calculate the Fisher Information Matrix.  This improves the numerical stability for the inversion operation to obtain the covariance matrix, and proposes MCMC jumps in the eigen-space of the Fisher Information matrix. Both methods implement a double-stage annealing heat treatment in order to effectively explore the parameter space at the beginning of the chain.

The rest of the paper is laid out as follows.  In section~\ref{lisapf} we provide details on the LISA Pathfinder dynamics and on the experiment for the calibration of parameters of the dynamics along the principal measurement axis. In section~\ref{mcmc} we develop the mathematical basis of our algorithm, while in section~\ref{results} we report the results of the application of our methods to LISA Pathfinder and we analyze the two methods in terms of acceptance rate, convergence to the equilibrium distribution and error on the test mass noise estimation caused by the uncertainty on the estimated parameters.

\section{A test case for LISA Pathfinder mission}
\label{lisapf}

LISA Pathfinder is a controlled three-body system composed of two test masses and the enclosing spacecraft. One test mass is free-falling along the principal measurement axis and it is used as a reference for the drag-free controller of the spacecraft. The second test mass is actuated at very low frequencies (below $1$ mHz) in order to follow the free falling test mass. This actuation scheme provides a measurement bandwidth of $0.5 \leq f \leq 100$ mHz in which both the test masses can be considered to be effectively free-falling. The system has two output channels along the principal measurement axis, which sense the displacement of the spacecraft relative to the free-falling test mass and the relative displacement between the individual test masses. We focus our attention on the principal measurement axis dynamics for which we can build a model that contains at least the five key parameters reported in Table~\ref{tbl:table1}~\cite{Congedo2012} .

\begingroup
\squeezetable
\begin{table*}[h]
  \caption{Parameter descriptions and the simulation and parameter range values used in the simulation and in the MCMC search. `Simulation values' are the values used to produce the set of synthetic data, they represent the `true' values for the parameters in our test. The `search range' defines the extent of the uniform prior distribution used in the MCMC.
\label{tbl:table1}}
\begin{ruledtabular}
\begin{tabular}{ c  p{5cm}  c  c  }
Name & Description & Simulation Value & Search Range \\
\hline
\hline
 
 $A_{df}$ & Actuation gain of the drag-free control loop acting on spacecraft thrusters & $1.05$ & $[0.5, 1.5]$ \\
 
 $A_{lfs}$ & Actuation gain of the electrostatic suspension actuating the second test mass at low frequency & $1.05$ & $[0.5, 1.5]$ \\
  
$S_{\Delta 1}$ & Interferometer cross-talk coefficient coupling the first channel with the differential channel & $1\times10^{-5}$ & $[-1\times10^{-3}, 1\times10^{3}]$ \\
   
 $\omega_1^2$ & Stiffness coupling the motion of the spacecraft to the motion of the first test mass & $1.4\times10^{-6} \, \textnormal{s}^{-2}$ & $[1\times10^{-7}, 1\times10^{-5}] \, \textnormal{s}^{-2}$ \\
    
 $\omega_2^2$ & Stiffness coupling the motion of the spacecraft to the motion of the second test mass & $2.2\times10^{-6} \, \textnormal{s}^{-2}$ & $[1\times10^{-7}, 1\times10^{-5}] \, \textnormal{s}^{-2}$ \\
 
\end{tabular}
\end{ruledtabular}
\end{table*}
\endgroup

We simulated two classical experiments \cite{Congedo2012} for LISA Pathfinder: one stimulating the motion of the spacecraft with respect to the free-falling test mass and the other forcing a movement of the actuated test mass with respect to the free-falling mass. 
The signal in the first experiment is obtained by forcing the drag-free controller of the spacecraft to change sinusoidally its set-point. This sends a command to the thrusters that produces a sinusoidal movement of the spacecraft with respect to the free-falling test mass. In the second experiment we force a sinusoidal bias in the set-point of the system that controls the low frequency actuation of the second test mass. As a consequence a sinusoidal voltage is applied by the actuators to the second test mass and a movement with respect to the free-falling test mass is induced. 
For the generation of our synthetic signals we adopted the standard scheme in which the output of the instrument can be written as $s(t) = h(t) + n(t)$, where $h(t)$ is the time-dependent noiseless output of the LISA Pathfinder model for the dynamics (the model is described in Ref~\cite{Congedo2012}) and $n(t)$ is an additive noise contribution generated in accordance to the procedure described in Ref~\cite{Ferraioli2010}. The noise $n(t)$ is a Gaussian distributed correlated noise that reproduces the current knowledge of the LISA Pathfinder noise budget~\cite{Antonucci2011}. This noise is principally composed of thruster noise in the channel sensing the displacement of the spacecraft relative to the free falling test mass (first channel) and of capacitive actuation noise on the channel sensing the relative displacement between the test masses (differential channel). These two channels produce a displacement signal at the output that needs to be converted into force-per-unit-mass for the analysis of the noise affecting the test masses. The conversion is based on a model of the spacecraft dynamics that is written in terms of the parameters described in Table~\ref{tbl:table1}. The calibration of the  parameters of such a model therefore has a direct impact on the LISA Pathfinder test-mass noise budget, which is the main scientific target of the LISA Pathfinder mission.

\begin{figure}[t]
\vspace{5mm}
\begin{center}
\epsfig{file=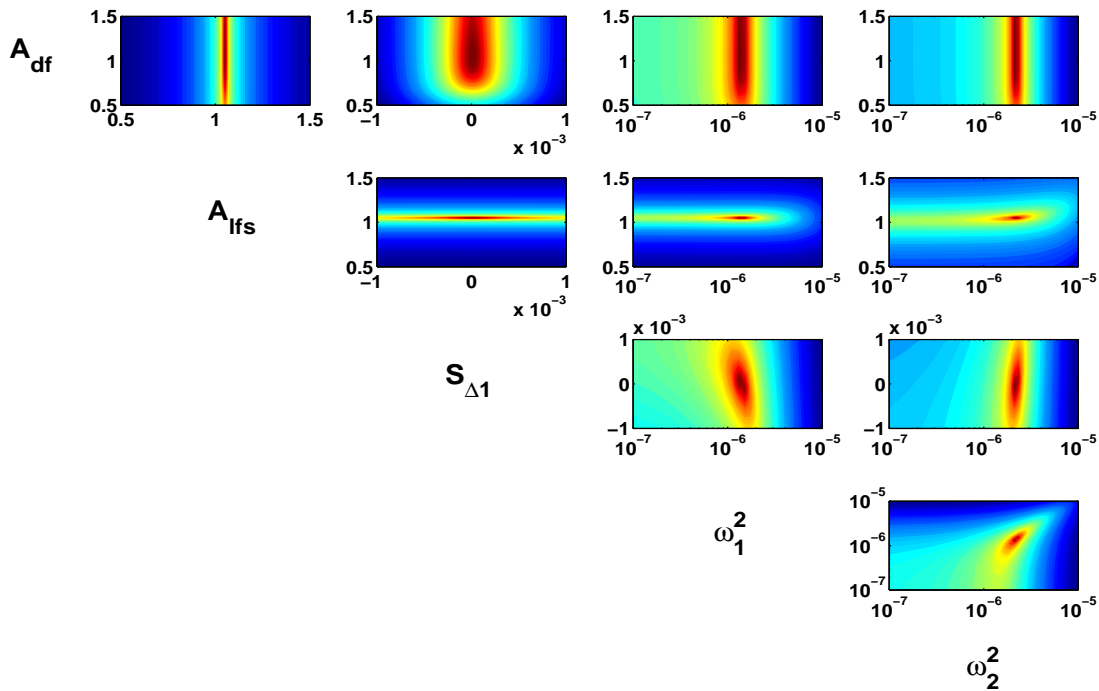, width=6.5in, height=4.2in}
\end{center}
\caption{2D slices in the 5D likelihood surface for the current parameter set.  Note that some of the axes are in log-scale while others are in a linear scale.  We clearly see that over the parameter ranges, the surface contains a single global solution.}
\label{fig:llsurf}
\end{figure}

In order to better understand our parameter estimation problem, we calculated the 5D likelihood surface over a uniform grid of points spanning the parameters ranges defined in Table~\ref{tbl:table1}. 2D slices of the likelihood surface are presented in Figure~(\ref{fig:llsurf}) where we clearly see that there is one global solution over the `search range'. The `search range' for the parameters is defined from the prior knowledge of the system that principally comes from construction constraints and requirements.

\section{Bayesian analysis and Markov Chain Monte Carlo} 
\label{mcmc}
Bayesian analysis has become a very useful tool in the field of gravitational waves~\cite{}, and has also been applied to LISA Pathfinder data analysis~\cite{Nofrarias2010}.  One of the advantages of Bayesian analysis is that it provides a direct method of calculating the marginalised posterior density function (PDF) for each of the parameters via Bayes' theorem
\begin{equation}
p(\vec{\lambda}|s) = \frac{\pi(\vec{\lambda})p(s|\vec{\lambda})}{p(s)},
\end{equation}
where $p(\vec{\lambda}|s)$ is the posterior density function (PDF) for the solution $\vec{\lambda}$ given the data $s(t)$, $\pi(\vec{\lambda})$ denotes the prior probability of the parameters and $p(s|\vec{\lambda})$ is the likelihood, which we will define below.  The constant in the numerator is the Bayesian evidence and is defined by
\begin{equation}
p(s) = \int d\vec{\lambda}\,\pi(\vec{\lambda})p(s|\vec{\lambda}),
\end{equation}
which as can be seen in independent of the solution set $\vec{\lambda}$ and is unimportant for our study.

While previous works have used sophisticated variants of the MCMC to search over a wide parameter space, in this study, as we can see from Figure~(\ref{fig:llsurf}), we are dealing with a simplified surface.  Even starting at the boundaries of our parameter priors, we can see that the surface contains a unimodal solution.  Because of this we can use a straightforward MCMC.  Starting with a detector output $s(t) = h(t) + n(t)$, where $h(t)$ is the induced signal and $n(t)$ is the noise in the detector, we generate a template $h(t;\vec{\lambda}_i)$  by choosing a random starting point in the parameter space $\vec{\lambda}_i$, and then draw from a proposal distribution in order to propose a jump to another point in the space $\vec{\lambda}_{i+1}$.  To compare both points, we evaluate the Metropolis-Hastings ratio
\begin{equation}
H = \frac{\pi(\vec{\lambda}_{i+1})p(s|\vec{\lambda}_{i+1})q(\vec{\lambda}_i|\vec{\lambda}_{i+1})}{\pi(\vec{\lambda}_i)p(s|\vec{\lambda}_i)q(\vec{\lambda}_{i+1}|\vec{\lambda}_i)},
\end{equation}
where we now define $p(s|\vec{\lambda})$ as the likelihood by
\begin{equation}\label{eqn:likelihood}
p(s|\vec{\lambda})\equiv{\mathcal L}\left(\vec{\lambda}\right) = C\,e^{-\left<s-h\left(\vec{\lambda}\right)|s-h\left(\vec{\lambda}\right)\right>/2},  
\end{equation}
where $C$ is a normalization constant.  We are unconcerned by this constant as we ultimately are interested in the likelihood ratio between the original and proposed points, hence cancelling the constant.  The quantity $q(\vec{\lambda}|\vec{y})$ is the transition kernel or proposal distribution used for jumping from $\vec{\lambda}_i$ to $\vec{\lambda}_{i+1}$.  Once a jump is proposed, it is then accepted with probability $\alpha = min(1,H)$, otherwise the chain stays at $\vec{\lambda}_i$.   While we are sure that the MCMC will eventually converge to the global solution, the timescale of this convergence is unclear.  So, while we can never be sure that our starting values are close to the true values, it can take a long time for the chain to ``walk" to the global maximum.  A common way of decreasing this travel time is to heat the likelihood surface using a technique called simulated annealing.  This has the effect of lowering and fattening structures on the surface.  We will describe this method in more detail at a later stage.

For the LISA Pathfinder experiments, we expect to have an extremely high signal-to-noise ratio (SNR) ($\sim 10^5 - 10^6$).  In this case the detector output has the form $s(t)\approx h(t)$.  For this high SNR limit, it is expected that the errors in the parameter estimation will have a Gaussian probability distribution given by
\begin{equation}
p\left(\Delta\vec{\lambda}\right)=\sqrt{\frac{\Gamma}{2\pi}}e^{-\frac{1}{2}\Gamma_{\mu\nu}\Delta \lambda^{\mu}\Delta \lambda^{\nu}},
\end{equation}
where $\Gamma_{\mu\nu}$ is the Fisher information matrix (FIM)
\begin{equation}
\Gamma_{\mu\nu} = \left<\frac{\partial h}{\partial \lambda^{\mu}}\left|\frac{\partial h}{\partial \lambda^{\nu}}\right.\right>,
\end{equation}
and $\Gamma = det\left(\Gamma_{\mu\nu}\right)$.  The angular brackets above denote the noise-weighted inner product for two real functions $s(t)$ and $h(t)$
\begin{equation}\label{eqn:scalarprod}
\left<h\left|s\right.\right> =2\int_{0}^{\infty}\frac{df}{S_{n}(f)}\,\left[ \tilde{h}(f)\tilde{s}^{*}(f) +  \tilde{h}^{*}(f)\tilde{s}(f) \right],
\end{equation}
where
\begin{equation}
\tilde{h}(f) = \int_{-\infty}^{\infty}\, dt\, h(t)e^{2\pi\imath ft}
\end{equation}
is the Fourier transform of the time domain waveform $h(t)$ and an asterisk denotes a complex conjugate.  The quantity $S_{n}(f)$ is the one-sided noise spectral density of the detector.  For the LISA Pathfinder project, there is no closed-form solution to the Fisher matrix calculation as it is not possible to take the derivatives of the template with respect to the parameters, nor evaluate the integral analytically.  Therefore, the derivatives in the inner product need to be calculate numerically using a one-sided difference algorithm.  Also, as the templates are expensive to generate (on the order of seconds), we update the FIM calculation every 100 iterations during the annealing scheme and as we believe that the FIM should remain almost constant at the global solution, every 5000 iterations during the MCMC.

While we have seen that the likelihood surface is highly simplified compared to other gravitational wave problems, in that there exists a single mode solution, our goal here is to present an algorithm that can be started from any point in the parameter space.    In order to accelerate the convergence of the chain to the global peak, we use a mixture of annealing methods as presented in~\cite{Cornish2007a,Cornish2007b}.  This first means re-writing the likelihood as
\begin{equation}
{\mathcal L}\left(\vec{\lambda}\right) = C\,\exp\left(-\frac{1}{2\beta}\left<s-h\left(\vec{\lambda}\right)|s-h\left(\vec{\lambda}\right)\right>\right),
\end{equation}
where we introduce a heat scaling $\beta$.  The first annealing scheme used is called thermostated annealing.  While the SNR for the experiments are large, and even though there is no chance of getting stuck on a secondary solution in this particular problem, it is difficult to choose the initial annealing temperature by hand.  If the temperature is too high, all structures on the likelihood surface are completely flattened and the MCMC aimlessly explores a flat surface.  Therefore, as it is again costly to generate templates, we do not want to waste computer cycles by choosing the initial temperature too high.  On the other hand, in the case of a multimodal solution, if the temperature is too low, the chain can get stuck on a secondary maximum and never reach the global solution.  Therefore, we allow the algorithm to chose its own initial annealing temperature according to   
\begin{equation}
\beta = \left\{ \begin{array}{ll} 1 & 0\leq SNR\leq SNR_T \\ \\ \left(\frac{SNR}{SNR_T}\right)^{2} & SNR > SNR_T  \end{array}\right. ,
\end{equation}
which ensures that once we attain a SNR of greater than a threshold SNR, $SNR_T$, the effective SNR never exceeds this value.   In our algorithm we run the thermostated annealing for the first 2000 iterations. For this problem, we choose our threshold SNR as follows.  As previously stated, the SNRs involved in this problem are very large.  If we take the boundary values for the parameters, as given in Table~\ref{tbl:table1}, the SNR at the boundary is $1.473\times10^6$, while at the injected values we have a SNR of $1.486\times10^6$.  To choose the threshold SNR, we arbitrarily demand an initial value of $\beta=500$.   Solving for the threshold SNR in the above expression then provides a threshold value of $SNR_T=6.58\times10^5$.  After this we use a second annealing scheme, which is standard simulated annealing, to cool the surface down.  To do this we use an annealing of the form
\begin{equation}
\beta = \left\{ \begin{array}{ll} 10^{\xi\left(1-\frac{i}{T_{c}}\right)} & 0\leq i\leq T_{c} \\ \\ 1 & i > T_{c}  \end{array}\right. ,
\end{equation}
where $i$ represents the chain index, and $T_c$ represents the number of chain points during the cooling schedule.   For this study,  $\xi=log_{10}(\beta_{max}^{TA})$ is the initial simulated annealing heat-index defined by the maximum temperature at the end of the thermostated annealing phase $\beta_{max}^{TA}$.  Again, as there is no risk of getting stuck on secondary maxima, we cool down the surface much quicker that we normally would for other problems.  Here we set the cooling schedule equal to $T_c=3000$ iterations.

One of the things that we wanted to test in this study was the difference between moving in eigen-directions and coordinate directions when we update the parameters.  In Ref~\cite{Cornish2007a,Cornish2007b} the proposed updates were done in the following manner.  Once the Fisher matrix is calculated, we find the eigenvalues $E_{\mu}$ and eigenvectors $V_{\mu\nu}$.  We then update the parameters in the eigen-directions according to 
\begin{equation}
\lambda^{\mu}\rightarrow \lambda^{\mu}+\left(\sqrt{\frac{\beta}{D}}\right)\sum_{\nu=1}^{5}\frac{V_{\mu\nu}}{\sqrt{E_{\nu}}} n_{\nu}
\end{equation}
where $D$ is the dimensionality of the problem and $n_{\nu} = {\mathcal N}_{\nu}(0,1)$ is a vector of normal distributed random numbers with zero mean and unit variance.  The factor of $1/\sqrt{D}$ ensures that the total jump in the parameter space has a length of $\sim 1\,\sigma$, while the heat factor $\beta$ ensures that while the temperature is high, we take larger steps in the chain and thus faster explore the parameter space 

To move in coordinate directions, we first invert the Fisher matrix to obtain the variance-covariance matrix, i.e.
\begin{equation}
{\mathbf C} \equiv C_{\mu\nu} = \left(\Gamma_{\mu\nu}\right)^{-1}.
\end{equation}
As ${\mathbf C}$ is a real symmetric matrix, we then use a Cholesky decomposition to write the covariance matrix as ${\mathbf C} = {\mathbf L}{\mathbf L}^*$, where ${\mathbf L}^*$ is the Hermitian conjugate or adjoint matrix of ${\mathbf L}$, and ${\mathbf L}$ is a unique lower triangular matrix.  We should note here that for a real symmetric matrix the Hermitian conjugate is equivalent to the matrix transpose.  We then generate a vector of zero mean, unit variance random Gaussian values $\vec{n}=n_{\nu}$, and we update the parameter values according to
\begin{equation}
\lambda^{\mu}\rightarrow \lambda^{\mu}+\sqrt{\beta}\left({\mathbf L}.\vec{n}\right),
\end{equation}
where the update is along the coordinate directions for the parameters.
\begin{figure}[t]
\vspace{5mm}
\begin{center}
\epsfig{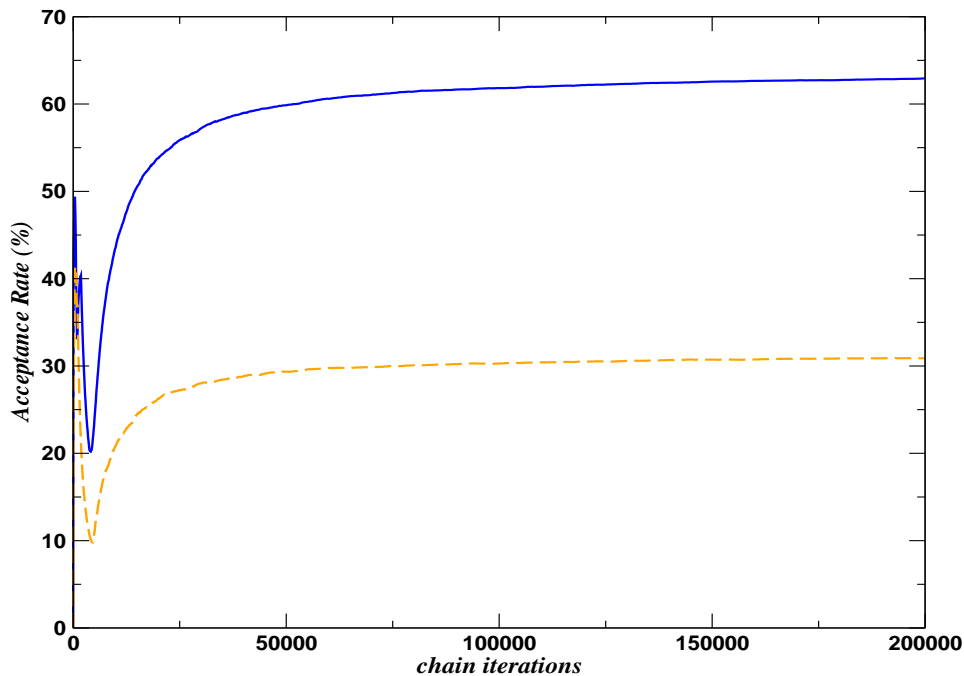}
\end{center}
\caption{The instantaneous acceptance rate for two Markov chains moving in eigen-directions (blue-solid) and coordinate directions (orange-dashed).  We observe a factor 
of two improvement by moving in eigen-directions, as opposed to moving in coordinate directions.}
\label{fig:accrate}
\end{figure}

In the analysis conducted in this article, we combine the information provided by two LISA Pathfinder output channels and two separate experiments. The inner product in Equation (\ref{eqn:scalarprod}) can be generalized in a straightforward way to the case of multiple output channels by

\begin{equation}\label{eqn:mchinner}
\left<{\mathbf s}-{\mathbf h}(\vec{\lambda})\left|{\mathbf s}-{\mathbf h}(\vec{\lambda})\right.\right> = 2 \int\limits_0^{\infty} 
 \left[ \tilde{s}_1 - \tilde{h}_1(\vec{\lambda}), \ldots, \tilde{s}_{Nc} - \tilde{h}_{Nc}(\vec{\lambda}) \right] \Sigma^{-1}_n(f) 
 \begin{bmatrix}
\tilde{s}_1 - \tilde{h}_1(\vec{\lambda}) \\
\vdots \\
\tilde{s}_{Nc} - \tilde{h}_{Nc}(\vec{\lambda}) \end{bmatrix}^{\ast} \,\mathrm{d}f + c. c.
 \end{equation}

Here $\Sigma_n(f)$ is the cross-spectral matrix, which contains the power spectral density at frequency $f$ on the diagonal and the cross-spectral density on the off-diagonal terms, $Nc$ is the number of output channels. 
Since we assume that the outputs of two distinct experiments are independent, the total likelihood for a number $N_e$ of experiments is just the sum of the likelihoods from each experiment, i.e.

\begin{equation}
\mathcal{L}_{tot} = \sum\limits_{i=1}^{N_e} \mathcal{L}_i.
\end{equation}

\section{Results}\label{results}
Our study involves two different types of MCMC.  For the movement in eigen-directions, we choose to work with the parameter set $\lambda^{\mu} = \left\{A_{df}, A_{lfs}, S_{\Delta 1}, \ln\,\omega_{1}^{2}, \ln\,\omega_{2}^{2} \right\}$.  By working in the natural logarithm of the parameters $\left\{\omega_{1}^{2}, \omega_{2}^{2}\right\}$ we lower the condition number of the Fisher matrix, making it numerically more stable for inversion purposes.  This matrix is still not as stable as we would like due to the small values of the parameter $S_{\Delta1}$.  However, as this parameter can also have negative values we cannot use its logarithm.  For the coordinate direction chains, we use the physical parameter values.   Each chain was run for $2\times10^5$ iterations, with 5,000 iterations of annealing.  To confirm the relative simplicity of the problem, starting with injected values of $\{1.885, 2.0399, -0.00746, 9.135\times10^{-6}, 6.327\times10^{-6}\}$, both chains converged to the global maximum within 3000 iterations.  However, just to be safe, we will label the first 8,000 chain points as ``burn-in".  These points will be rejected when it comes to constructing a statistical analysis from the chains.  These initial values correspond to a separation of approximately $\{10^3, 10^5, 10^4, 10^4, 10^4\}\sigma$ from the injected values, where $\sigma$ here is the standard deviation calculated from the chain and presented later in Table~\ref{tab:stat}.  In Figure~(\ref{fig:accrate}) we plot the instantaneous acceptance rates for both the eigen-direction chain (blue-solid) and the coordinate direction chain (orange-dashed).  We can see that we gain roughly a factor of two by using jumps in eigen-directions rather than coordinate directions with the chains having final acceptance rates of $63\%$ and $31\%$ respectively.  This means that to independently sample the same posterior density, the coordinate direction chain would need to be twice as long as the eigen-direction chain.

\begin{figure}[t]
\vspace{5mm}
\begin{center}
\epsfig{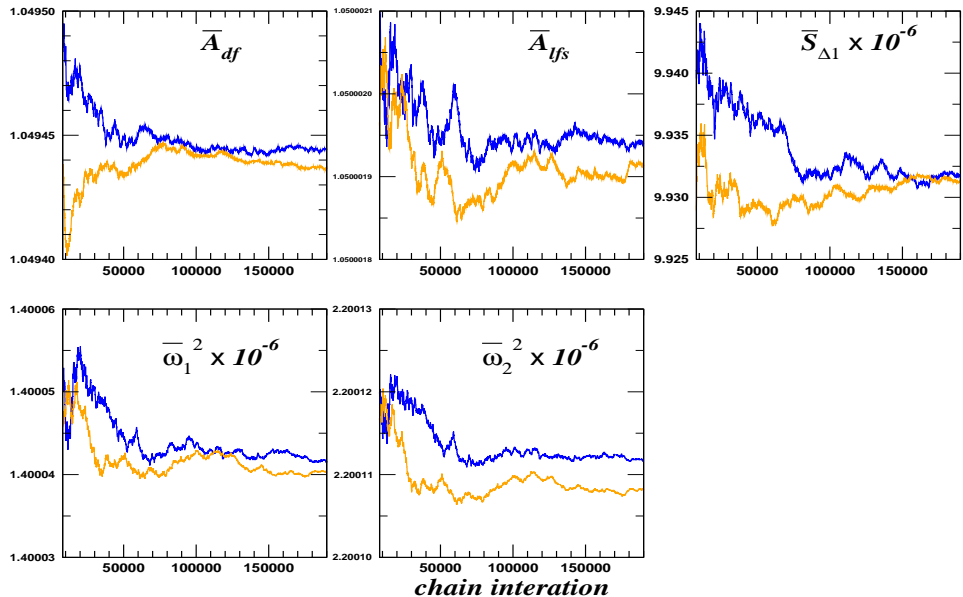}
\end{center}
\caption{The instantaneous parameter means for two Markov chains moving in eigen-directions (blue) and coordinate directions (orange). The two methods show similar rate of convergence and despite we are dealing with an unimodal problem it takes $10^5$ samples to have a reasonable convergence.}
\label{fig:means}
\end{figure}

\begin{figure}[h]
\vspace{5mm}
\begin{center}
\epsfig{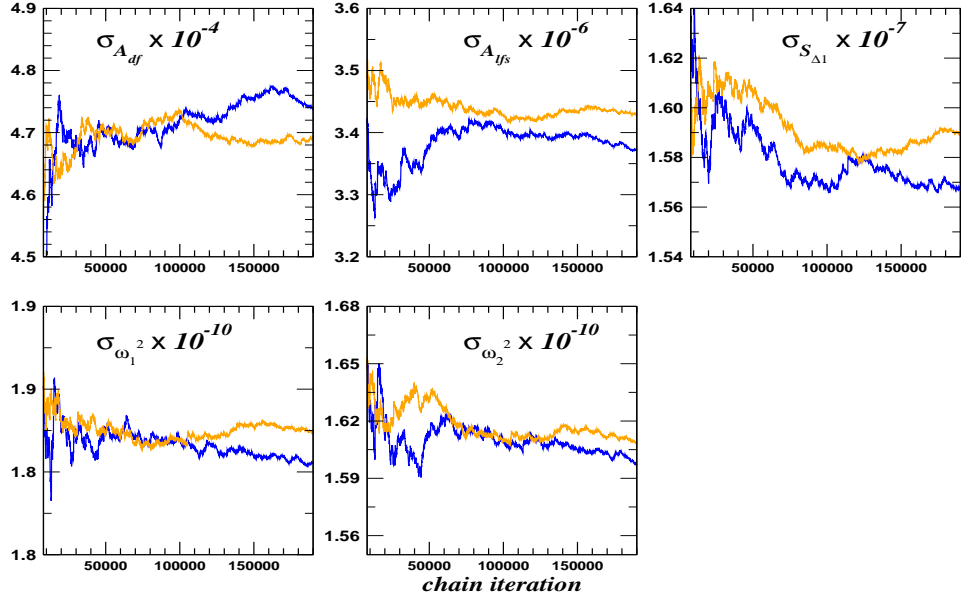}
\end{center}
\caption{The instantaneous parameter standard deviations for two Markov chains moving in eigen-directions (blue) and coordinate directions (orange). }
\label{fig:sds}
\end{figure}

In Figures~(\ref{fig:means}) and~(\ref{fig:sds}) we plot the convergence of the means and standard deviations of each parameter from the two MCMC methods.  We can see from comparing both methods that they have a similar rate of convergence.  We should point out here that we would not expect both chains to find exactly the same value.  If we were to run $N$ chains, we would expect the chains to produce $N$ values all roughly within $\pm1\sigma$ of the true value.  Our criterion of evaluation here is how quickly the chains flatten with no large fluctuations.  Although after $2\times10^5$ iterations, it is still difficult to prefer one method over the other.  This is because of the fact that even though we are dealing with a simplified problem, it still takes $10^5$ iterations for the Metropolis-Hastings algorithm to begin to converge.  This suggests that for full convergence, the chains would need to be run for longer.  However, it is still clear that we are finding the global peak very quickly.

\begin{figure}[t]
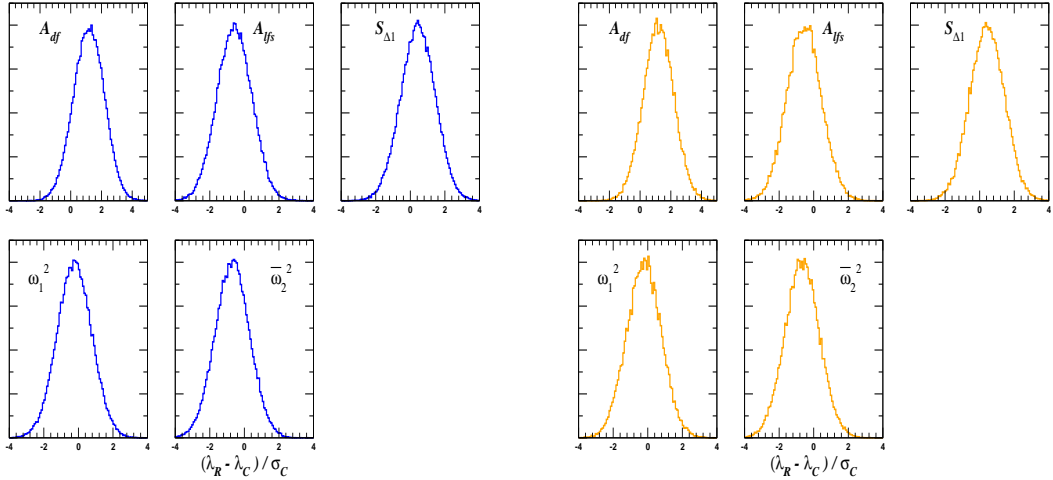

  \vspace{5pt}

  \centerline{\hbox{ \hspace{0.0in} 
   \epsfig{file=Fisher_Histograms.eps, width=2.5in, height=2.5in}
    \hspace{1cm}
    \epsfig{file=MVG_Histograms.eps, width=2.5in, height=2.5in}
    }
  }

  \caption{ This Figure displays the marginalised 1D histograms from the chains moving in eigen-directions (left) and coordinate directions (right).  The x-axis
origin is set at the true value of the parameters and the offset is the scaled difference between the true value $\lambda_R$ and the chain value $\lambda_C$,
divided by the standard deviation of the chain $\sigma_C$. Both chains give distributions very close to the true values but from a closer observation the coordinate direction distributions have a higher kurtosis due to the lower acceptance rate.}
  \label{fig:hists}

\end{figure}

In Figure~(\ref{fig:hists}) we plot the marginalised, normalised histograms from the MCMC.  On the x-axis we plot the true value offset scale $(\lambda_R - \lambda_C)/\sigma_C$, where $\lambda_R$ is the true injected parameter value, $\lambda_C$ is the chain value and $\sigma_C$ is the standard deviation as calculated from the chain.  We can see that both chains give us distributions very close to the true values.  If we look closer, we can however see that the coordinate direction distributions have a higher kurtosis due to the lower acceptance rate.   In all cases, the distributions are almost Gaussian with a skewness of $\sim10^{-2}$.   While there may seem to be no reason to prefer one method over the other, we will focus on the results from the eigen-direction chain due to its higher acceptance rate.  We present the results in full in Table~\ref{tab:stat}, but we quote here that the chain returned mean values for each parameter that were within 
$\{1.17, -0.58, 0.44, -0.23, -0.69\}\sigma$ of the true answer.

\begin{table}[t]
\caption{Simulated and recovered parameter values from the eigen-direction MCMC.  The final column gives the standard deviation scaled offset from the true value.}
\label{tab:stat}
\begin{tabular}{ c  c  c  c  }
Simulation Value & Mean & Standard Deviation & $(\lambda_R - \lambda_C)/\sigma_C$  \\
\hline
\hline
 
 1.05 &  1.049445 & $4.7356\times10^{-4}$ & 1.17 \\
 
 1.05 &  1.050002 & $3.3727\times10^{-6}$ & -0.58 \\
  
$10^{-5}$  & $9.9314\times10^{-6}$ & $1.5669\times10^{-7}$ & 0.44 \\
   
 $1.4\times10^{-6}$  & $1.400041\times10^{-6}$ & $1.8048\times10^{-10}$ & -0.23 \\
    
 $2.2\times10^{-6}$  & $2.200111\times10^{-6}$ & $1.5974\times10^{-10}$ & -0.69 \\
 
\end{tabular}
\end{table}

\begin{table}[t]
\caption{Test mass noise response to parameters variations. The response of the noise as a function of each parameter is linear $n(\nu) = A \nu + B$, therefore we obtained the slope $A$ by a linear fit to the data. The noise variation $\Delta n$ at $1 \sigma$ is obtained multiplying $A$ by the sample standard deviation of the parameters obtained from the chain.}
\label{tab:noisesen}
\begin{tabular}{ c c  c  c  }
$\lambda$ &\,\,\,\,\,\,\,\,\, & $A$&  $\Delta n \textnormal{ at }  1 \sigma \,  [\textnormal{m} \textnormal{s}^{-2} \textnormal{Hz}^{-1/2}] $   \\
\hline
\hline
 
 $A_{df}$&  &  $1.9\times10^{-28}$ & $9\times10^{-32}$  \\
 
 $A_{lfs}$&  & $6.9\times10^{-15}$  & $2\times10^{-20}$  \\
  
$S_{\Delta1}$ &  & $3.5\times10^{-16}$ & $5\times10^{-23}$  \\
   
 $\omega_1^2$&  & $2\times10^{-12}$ & $4\times10^{-22}$  \\
    
 $\omega_2^2$ & & $1.2\times10^{-11}$ & $2\times10^{-21}$  \\
 
\end{tabular}
\end{table}
One of the aspects of the problem that we can also investigate from the MCMC is the correlations between parameters.  Using the chains, we calculate the variance-covariance matrix using
\begin{equation}
C_{\mu\nu} = \frac{1}{N-1}\sum_{i=1}^N\,\left(x_{\mu} - \overline{x}\right)\left(y_{\nu} - \overline{y}\right),
\end{equation}
where $N=192,000$ is the number of chain points (ignoring the 8,000 burn-in points at the beginning of the $2\times10^5$ iterations).  An overbar denotes the mean of the parameter as calculated from the chain.  Once we have the variance-covariance matrix, we can calculate the matrix of standard deviations and correlation coefficients according to 
\begin{equation}
D_{\mu\nu} = \left\{ \begin{array}{lll}
\sqrt{C_{\mu\nu}} & \,\,\,\,\, & \mu = \nu\\
C_{\mu\nu} / \sqrt{\sigma_{\mu}\sigma_{\nu}} & \,\,\,\,\, & \mu \neq \nu \end{array}\right. ,
\end{equation}
where $\sigma_{\mu}=\sqrt{C_{\mu\mu}}$ are the standard deviations of the marginalised chains for each parameter.  The numerical values for the matrix, where the order of the rows and columns correspond to $\left\{A_{df}, A_{lfs}, S_{\Delta 1}, \omega_{1}^{2}, \omega_{2}^{2} \right\}$ are
\begin{equation}
D_{\mu\nu} = \left( \begin{array}{rrrrr}
4.736\times10^{-4}  &  -9.921\times10^{-3} &  1.662\times10^{-3}  &   -6.891\times10^{-3} &   -1.547\times10^{-3}  \\
 & 3.373\times10^{-6} &    1.259\times10^{-2}  &   6.697\times10^{-1}  &   7.659\times10^{-1}  \\  
   &   &   1.567\times10^{-7}  &   -2.267\times10^{-1} &   8.097\times10^{-3}   \\ 
 &    &    &   1.805\times10^{-10}   &  8.789\times10^{-1}    \\
  &   &     &      &  1.597\times10^{-10}\end{array} \right).
\end{equation}
We can see that the parameter set is mostly uncorrelated.  However, we do see strong correlations between $A_{lfs}$ and the two parameters $(\omega_{1}^{2},\omega_{2}^{2})$, i.e. 0.67 and 0.77, and between $\omega_{1}^{2}$ and $\omega_{2}^{2}$ themselves, i.e. 0.88.

It is interesting here to quantify the accuracy of our result in terms of the LISA Pathfinder test mass noise budget. As discussed above, the parameters of the dynamics play an important role in the analysis of the test mass noise, since they are used for the calibration of the model that converts the displacement noise to force-per-unit-mass. On the basis of such a model we calculated the noise response to the parameters as reported in Table~\ref{tab:noisesen}. This data is obtained calculating the noise-per-unit-mass on an uniform grid of values for the parameters. The grid is centered around the simulation `true value' and extends over $10 \sigma$, where $\sigma$ is the sample standard deviation obtained by the chain. 
The response of the noise as a function of each parameter is linear $n(\nu) = A \nu + B$, therefore we obtained the slope $A$ reported in Table~\ref{tab:noisesen} by a linear fit to the data. The noise change $\Delta n$ at $1\sigma$ is obtained by multiplying $A$ by the values of $\sigma$ obtained by the chain.
As can be seen in Table~\ref{tab:noisesen}, even combining the results for all the five parameters we obtain $\Delta n \sim 2 \times 10^{-21} [\textnormal{ m} \textnormal{s}^{-2} \textnormal{Hz}^{-1/2}]$. Even assuming that we use 6 months data (the full mission duration) this value is three orders of magnitude lower than the expected experimental uncertainty ($2.3 \times 10^{-18} \, [\textnormal{m} \textnormal{s}^{-2} \textnormal{Hz}^{-1/2}]$).
For this calculation we calculated the power spectral density with the well known Welch's overlapped segment averaging method, using a $4$ sample Blackman-Harris window \cite{Harris1978} and $50\%$ segment overlapping. A reasonably long LISA Pathfinder noise run can last for $3$ days, and the current estimation for the noise level is $7.2 \times 10^{-15} \, [\textnormal{m} \textnormal{s}^{-2} \textnormal{Hz}^{-1/2}]$ at $1 \textnormal{ mHz}$ \cite{Antonucci2011}. Using this information in calculating the expected uncertainty, we obtain a value of $1.4 \times 10^{-16} \, [\textnormal{ m} \textnormal{s}^{-2} \textnormal{Hz}^{-1/2}]$. While if we assume a $6$ month long data run, we obtain an experimental uncertainty of $2.3 \times 10^{-18} \, [\textnormal{m} \textnormal{s}^{-2} \textnormal{Hz}^{-1/2}]$, where the uncertainty is simply calculated as the value for the power spectral density divided by the number of averaging segments.

\section{Conclusions}
\label{conclusions}
We have presented a study of two different MCMC methods for an experiment with LISA Pathfinder.  While both methods are based on the Metropolis-Hastings algorithm, and both use the Fisher matrix to generate the covariance matrix, they differ in their treatment of the parameters and the way in which they move in the parameter space.  In one case, we work with the physical parameter values and make MCMC jumps in the coordinate directions.  For the other chain, we generate a more numerically stable Fisher matrix by working in the logarithm of two of the five parameters.  This lowers the condition number of the matrix and makes it more stable during the inversion process to obtain the covariance matrix.  We demonstrate that while the experiment chosen for this study is quite simple, in that a singe global solution exists in the likelihood surface, we gain roughly a factor of two in acceptance rate by moving in eigen-directions rather than in coordinate directions.  Our chains both end up with parameter estimations that are very close to the true values.  This is to be expected due to the extremely high SNR of the signal.  However, as we are conducting a Bayesian inference, we are mostly interested in the full posterior distribution for each parameter.  In this case, due to the higher acceptance rate, it is clear that calculating a more stable Fisher matrix and moving in eigen-directions is an optimal choice.
The accuracy of the recovered parameters has been analyzed in terms of the effect they have on the force-per-unit-mass test mass noise estimate. We demonstrated that the error introduced by the parameters estimated from the MCMC chain is in any case more than three orders of magnitude less than the expected experimental uncertainty in the power spectral density.

\begin{acknowledgments}
This research was supported by the Centre National D'\'etudes Spatiales (CNES).
\end{acknowledgments}


\bibliography{FerraioliBib01}{} 

\end{document}